\documentclass[fleqn,usenatbib]{mnras}

\usepackage{newtxtext,newtxmath}
\usepackage{appendix}

\usepackage[T1]{fontenc}
\usepackage{comment}
\usepackage{multirow}
\DeclareRobustCommand{\VAN}[3]{#2}
\let\VANthebibliography\thebibliography
\def\thebibliography{\DeclareRobustCommand{\VAN}[3]{##3}\VANthebibliography}

\usepackage{graphicx}	
\usepackage{amsmath}	
\usepackage{siunitx}

\usepackage{makecell, caption, booktabs, multirow, siunitx}

\usepackage{lipsum, fancyvrb}

\title[The rocket effect in neutron stars]{The rocket effect mechanism in neutron stars in supernova remnants}

\author[V. Agalianou \& K.N. Gourgouliatos]{
V. Agalianou,$^{1}$\thanks{E-mail: v.agalianou@uoi.gr (VA)}
K. N. Gourgouliatos,$^{1}$ \thanks{E-mail: kngourg@upatras.gr (KNG)}\\
$^{1}$Department of Physics, University of Patras, Patras, Rio, 26504, Greece\\
}

\date{Accepted XXX. Received YYY; in original form ZZZ}

\pubyear{XXX}

\begin{document}

\label{firstpage}
\pagerange{\pageref{firstpage}--\pageref{lastpage}}
\maketitle

\begin{abstract}
While the dipole magnetic field axis of neutron stars is usually postulated to cross the star's centre, it may be displaced from this location, as it has been recently indicated in the millisecond pulsar J0030+0451. Under these conditions, the electromagnetic rocket effect may be activated, where the magnetic field exerts a net force, accelerating the star. This post-natal kick mechanism relies on asymmetric electromagnetic radiation from an off-centre dipole may be relevant to the high spatial velocities of pulsars $\sim 10^{3}$ $\rm{km/s}$. Here, we explore its impact in young pulsars associated with supernova remnants and we compare the observational data on characteristic quantities, such as the braking index and proper motion, with results obtained from the rocket effect. Using a Markov Chain Monte Carlo analysis, we explore the required conditions, for the initial spin periods and the distance between the magnetic axis and the star's center, so that the velocity kick due to the rocket effect approaches the present velocity. We find that the electromagnetic rocket effect can account for typical pulsar transverse velocities assuming an initial spin period of 3.8 $\rm{ms}$ and a dipole field whose distance from the centre of the star is approximately 7 $\rm{km}$. We also explore the influence of the rocket effect on the braking index of a neutron star, and we find that for the sample studied this impact is minimal. Finally, we apply the rocket effect model on the pulsars J0030+0451 and J0538+2817, which are likely candidates for this mechanism.

\end{abstract}

\begin{keywords}
stars: neutron--kicks--kinematics
\end{keywords}



\section{Introduction}
The velocities of neutron stars are in several occasions rather high, with average values reaching up to 200-500 $\rm{km/s}$  \citep{lyne1994high,cordes1998neutron,10.1093/mnras/289.3.592}, while in some cases they are exceeding $1000$ $\rm{km/s}$ \citep{chatterjee2005getting}. Recently,  Gaia Data Release 2 has provided precise parallax measurements, and the velocities of many pulsars have been estimated to belong in a wide range of speeds \citep{yang2021revisiting}. Since the progenitors of pulsars are believed to have smaller space velocities \citep{blaauw1985progenitors}, a physical explanation
on how some pulsars obtain high velocities, is the assumption that there was an effect of a kick mechanism on the star \citep{coleman2022kicks}, either at its birth \citep{2020MNRAS.494.3663I,2021MNRAS.508.3345I}, or at some following phase of the pulsar's life \citep{bailes1989origin}. 

Some of the main kick-models that have been proposed are the following: A velocity kick can be generated by hydrodynamic instabilities which occur during the core collapse of the proto-neutron star or during a supernova explosion \citep{10.1093/pasj/psz080,Burrows_1995, Herant1994InsideTS}. During the collapse of the star, uneven disturbances appear in its outer remaining core of iron, which are amplified by gravity. These enhanced disturbances lead to sharp propagation of shocks that have as a result the appearance of kick. \cite{janka1994neutron} performed 3-dimensional hydrodynamical simulations of the evolution of the perturbations as they are transported to the outer regions of the neutron stars. They found that these velocity kicks can hardly 
exceed 100 $\rm{km/s}$, since the convective blobs, which carry the radiated energy, do not cause significant anisotropies. The estimated timescales of hydrodynamic kicks is 0.1 $\rm{s}$, for a typical neutron star when the perturbations travel fast \citep{Burrows_1996}, and for this mechanism to be viable, a massive progenitor is required \citep{janka2022supernova}. For most kick mechanisms, a spin-velocity alignment is expected, which is a phenomenon that has been best studied for the cases of Vela and Crab pulsars but also for other pulsars \citep{10.1111/j.1365-2966.2005.09669.x}. This spin-kick velocity correlation can be studied with pulsar wind nebula observations or by using the linear polarization profile of radio emission \citep{wang2007spin}. For the hydrodynamic kick the spin-kick alignment is disputed if the timescale of the kick is smaller than the spin period of the pulsar \citep{2001ApJ...549.1111L}. 

An alternate mechanism that has been proposed is acceleration due to neutrino emission because of the presence of strong magnetic fields \citep{10.1093/mnras/staa261, wongwathanarat2013three, colpi2002formation}. Asymmetric neutrino flux can be generated in the outer regions of the star, where the neutrino distribution deviates from thermal equilibrium \citep{vilenkin1995parity}. \cite{arras1999can} found that the kick velocity induced by this mechanism is $50 B_{15} $  $\rm{km/s}$, where $B=10^{15}~B_{15}$   $\rm{G}$, is the local magnetic field strength. So this mechanism requires very strong magnetic fields to result in a significant acceleration of the neutron star. \cite{fryer2006effects} studied the effect of neutrino-driven
kicks on the supernova explosion mechanism, and found that they can increase significantly the energy of an asymmetric explosion, provided that the magnetic field strength is magnetar-level while the timescale for this kick mechanism is just a few seconds \citep{2001ApJ...549.1111L}. Moreover, some models propose kicks that result from the combination of the two last mechanisms, i.e., the kicks due to supernova explosions and the asymmetric neutrino emission \citep{stockinger2020three,  burrows2020overarching}. For the neutrino driven kick, spin-kick alignment is expected, as long as the spin period of the pulsar is small \citep{2001ApJ...549.1111L}. We summarize the main kick mechanisms and predictions in Table \ref{kicks}.

Another candidate mechanism that has been proposed to explain neutron star acceleration, is the electromagnetic kick or rocket effect mechanism, which is the main subject of this paper and was first introduced by \citet{Harrison1975AccelerationOP}. The electromagnetic kick mechanism  requires a strong magnetic field which is displaced from the center of the neutron star and may have multipolar components.  Due to this offset magnetic field, an asymmetric electromagnetic emission is expected, therefore a radiation reaction force in the direction of the spin axis  acts onto the star and eventually accelerates it.  For this model, the spin-kick alignment is expected. That is due to the fact, that the electromagnetic force that accelerates the star acts like a recoil, and appears because of the emission of radiation. So even though this emitted radiation is not aligned with the magnetic axis, it will be aligned with the spin axis. Recently, observations of the pulse shape of pulsar PSR J0030+0451, that were obtained by NICER observatory in X-rays and Fermi-LAT in gamma-rays \citep{miller2019psr, riley2019nicer}, revealed that the morphology of the star's hot spots and polar caps imply an non-centered dipole magnetic field which could have multipole components. As shown by \citet{Kalapotharakos_2021}, to interpret these morphologies, pulsar J0030+0451 may have a magnetic field which can be described by a variety of forms, which in general are consisted of an offset magnetic dipole moment and an offset quadrupole magnetic moment. Since this first direct observation of an off-centred magnetic field, the rocket effect mechanism should be taken into consideration as one viable candidate for the acceleration of neutron stars. 

\begin{table*}
\setlength{\tabcolsep}{2.5pt}
\renewcommand{\arraystretch}{3.4}
\caption{The main kick mechanisms for neutron stars.} 
\begin{tabular}{|c| c|c| c| c| c|} 
\hline 
\textbf{Mechanism}  & \textbf{Velocity (\rm{km/s})}  & \textbf{Alignment of kick-spin} & \textbf{Natal/Post-natal} & \textbf{Timescale}  & \textbf{Requirements}\\
\hline 
\textbf{Hydrodynamical} & $\sim 50 $ $B_{15}$ & Is not always expected & Natal  & $\sim 0.1 \rm{s} $  &  Massive progenitors with high densities. \\ 
\textbf{Neutrino Driven} & $\sim 100$ & Is expected  & Natal & $\sim 10$ $  \rm{s}$ &  low-mass progenitors and a magnetic field that exceeds $10^{15}$ $\rm{G}$.\\ 
\textbf{Rocket effect} & $\sim 1400 $ $R_{10}^2 P_{ms}^{-2}$ & Is expected & Post-natal & a few years & Initial spin period almost $\sim 1$ $\rm{ms}$ and weak gravitational emission.\\
\hline 
\end{tabular}
\label{kicks}
\end{table*}

In addition to the acceleration of neutron stars, the rocket effect mechanism predicts that the thrust imparted (the kick velocity vector) will be parallel to the neutron star's axis of rotation (spin-kick alignment). Recently, \cite{2021NatAs...5..788Y} presented the first case of three-dimensional alignment of the spin axis with the vector of the pulsar's spatial velocity. This work was based on observations by the Five-hundred-meter Aperture Spherical radio Telescope (FAST) \citep{NAN_2011}. By observing the polarization of the radiation emitted by the star, they determined both the transverse and radial velocity of the pulsar, but also showed that the rotational axis is approximately aligned with the velocity vector, forming a slight slope of almost 10 degrees. Such an observation further reinforces the relevance of the study of the rocket effect mechanism.

Beyond the context of neutron stars, off-centred dipole magnetic fields have been presented by \citet{kemp1970discovery} for white dwarfs. Recently \cite{https://doi.org/10.48550/arxiv.2301.04665}, modeled  the magnetic field of the white dwarf SDSS J1143+6615 as an offset dipole based on observational data. Moreover, research has been done on the electromagnetic fields that develop near the  surface of neutron stars. Specifically, \cite{10.1093/mnras/stt2300} showed that the Hall effect at the neutron star's crust produces quadrupolar and multipolar magnetic field components, with more complex behaviour appearing in three-dimensional studies \citep{gourgouliatos2016magnetic, gourgouliatos2017magnetic}. Apart from white dwarfs and neutron stars, there have been evidence of an offset and multipolar magnetic field to exist on planets too, as Uranus and Neptune \citep{1986Sci...233...85N, 1989Sci...246.1473N} and the magnetospheric structure of Jupiter \citet{1976JGR....81.3407K}. So, it appears, that multipolar and offset magnetic fields are likely to appear in several astrophysical objects even though assuming a centered dipole magnetic field simplifies the calculations and simulations that one can do.

In this paper we apply the rocket effect model on a sample population of neutron stars associated with supernova remnants (SNRs) and we export results about the values of the braking indices of the pulsars and their kick velocities. The structure of the paper is as follows: in Section \ref{2}, we describe the rocket effect as first proposed by \cite{Harrison1975AccelerationOP} and we present an alternate model which is time-depended and explicitly depends on the magnetic field strength. We also present the sample population of pulsars which are associated with supernova remnants and we focus on their initial spin period and braking index. Section \ref{3} contains our results for the velocity kicks and show the results of Markov Chain Monte Carlo (MCMC) analysis, where we also  estimate the range of values of the distance $s$ of the magnetic axis from the star's center and the initial spin period. In Section \ref{4f} we focus on two case studies, pulsars J0030+0451 and  J0538+2817 which are of special interest as the radiative properties of the former imply an off-centred dipole and the observations of the latter suggest spin-velocity alignment, and apply again the rocket effect model. We discuss our results and conclude in Section \ref{Discussion}.

\section{Methodology}\label{2}

\subsection{The Rocket Effect}

\begin{figure}
    \centering
   \includegraphics[width=0.97\linewidth]{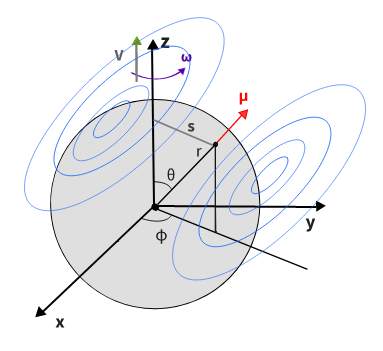}
    \caption{Geometry of the off-centred dipole for our model. The magnetic dipole moment \pmb{$\mu$} is displaced a distance $s$ from the spin axis ($z$) and has an oblique angle $\theta$. Also \pmb{$\omega$} is the angular velocity of the neutron star's rotation while \pmb{$V$} is the kick velocity vector. The blue ellipsoid lines indicate the magnetic field lines.}
    \label{fig:1}
\end{figure}

The electromagnetic kick or rocket effect model was first introduced by \citet{Harrison1975AccelerationOP} who suggested that a pulsar can be accelerated due to emission of asymmetric electromagnetic radiation. This asymmetric radiation can result from an oblique dipole moment displaced from the center of the star, depicted in Figure \ref{fig:1}.
According to \cite{Harrison1975AccelerationOP} and including \cite{2001ApJ...549.1111L} correction, neutron stars are expected to develop kick velocities:
\begin{equation}
    V_{kick}^{HT} \approx 140 \hspace{0.1cm} R_{10}^2 \hspace{0.1cm} \left(\frac{s}{10km}\right)\hspace{0.1 cm} \left(\frac{f_i}{1 kHz}\right)^3\left[ 1-\left(\frac{f}{f_i}\right)^3\right] \rm{km \hspace{0.1cm} s^{-1}},
    \label{eq1}
\end{equation}
where $R_{10}$ is the radius of the neutron star in units of 10 $\rm{km}$ and $f$, $f_i$ are the current and the initial rotational frequency respectively, and $s$ is the distance of the magnetic axis from the spin axis. It seems that for this model to be efficient and able to explain the current pulsar velocities we must consider a very small initial spin period. Therefore we expect this mechanism to appear in young millisecond pulsars with strong magnetic fields, which are usually still associated with supernova remnants.

\subsection{The time dependence of the kick - An alternate model }
\label{sec:alternative}
 
In this section we  further develop the electromagnetic kick mechanism to account for the spin-down of the star and the initial period. Thus we assume that the acceleration of the neutron star is only due to the force $\bf F$ given by equation \ref{F} (see appendix). We further assume that the  intensity and direction of the magnetic field do not evolve with time. Moreover, we include the  temporal evolution of the mechanism, since time is a parameter that has not been taken into account in the initial rocket effect model. Starting from Newton's second law:
\begin{equation}\label{new}
   {\bf F} = m \frac{d {\bf u}}{dt} \Rightarrow {\bf u}= \frac{1}{m} \int {{\bf F}\,dt},
\end{equation}
where F is given by the following equation:
\begin{equation*}
    { F}=\frac{8 \omega^5 \mu_z^2 s}{15 c^5}\,,
\end{equation*}
where the angular frequency is \textit{$\omega= 2\pi/ P$}, with $P$ being the pulsar period. We can evaluate the pulsar period as a function of time, as follows. Assuming that the strength of the magnetic field do not change, then due to the fact that $B^2 \propto P \dot{P} $, we have $P \dot{P}=$const., then:

\begin{equation*}
    P \dot{P}= P \frac{dP}{dt} \Rightarrow P \dot{P} dt = P dP \Rightarrow  \int_{0}^{t} { P \dot{P} dt}=  \int_{P_{in}}^{P} {P dP} \Rightarrow
\end{equation*}

\begin{equation}\label{tau_c}
    t=\frac{P^2-P_{in}^2}{2 P{\dot P}} \Rightarrow P(t)= \sqrt{2 \dot P P t+ P_{in}^2}\,.
\end{equation}
This provides the pulsar period as a function of time. Substituting the current period gives $\tau_c$,  the characteristic age of the pulsar. $\dot P$ is the period derivative and \textit{$P_{in}$} the initial spin period of the pulsar. To use equation \ref{tau_c}, we must assume that the braking index is equal to 3 and that the magnetic field and the moment of inertia do not change significantly with time. As long as we have the time dependency of the spin period, we can integrate from $t=0$, for the birth of the pulsar, until the age of the supernova remnant $\tau_{SNR}$ which is associated with the pulsar: 
\begin{equation}
    V_{kick}^{AL}= \frac{8 (2 \pi)^5 \mu_z^2 s^2}{15 m c^5} \int_{0}^{\tau_{SNR}} {\frac{1}{P(t)^{5}}\,dt},
\end{equation}  

\begin{equation}\label{v_AL_time}
        V_{kick}^{AL}= \frac{8 (2 \pi)^5 \mu_z^2 s^2}{15 m c^5} \int_{0}^{\tau_{SNR}} {\frac{1}{\left(2 \dot P P t+ P_{in}^2\right)^{5/2}}\,dt},
\end{equation}

\begin{equation}\label{10}
    V_{kick}^{AL}= \frac{8 (2 \pi)^5 \mu_z^2 s}{15 m c^5} \left[ -\frac{1}{3 P \dot{P}\left(2 P \dot{P} \tau_{SNR} + P_{in}^2\right)^{3/2}} + \frac{1}{3 P \dot{P} P_{in}^3} \right].
\end{equation}

Equation \ref{10} gives our alternate model for the kick velocity which depends on several parameters such as the magnetic dipole moment $\mu_z$, which we consider to be \textit{$\mu_z= BR_{*}^3$}.  Here $R_{*}$ is the radius of the neutron star. The main difference between equation \ref{eq1} and \ref{10} is the time dependence of the kick-velocity, and the fact that it results to different values for the kick velocity, as shown below. Having presented the two models from which we can derive the kick velocity, next we present the population of pulsars for which we apply these two models. 

\subsection{Sample of pulsars}

\begin{figure}
\hspace{-1.2cm}
 \includegraphics[width=1.1\linewidth]{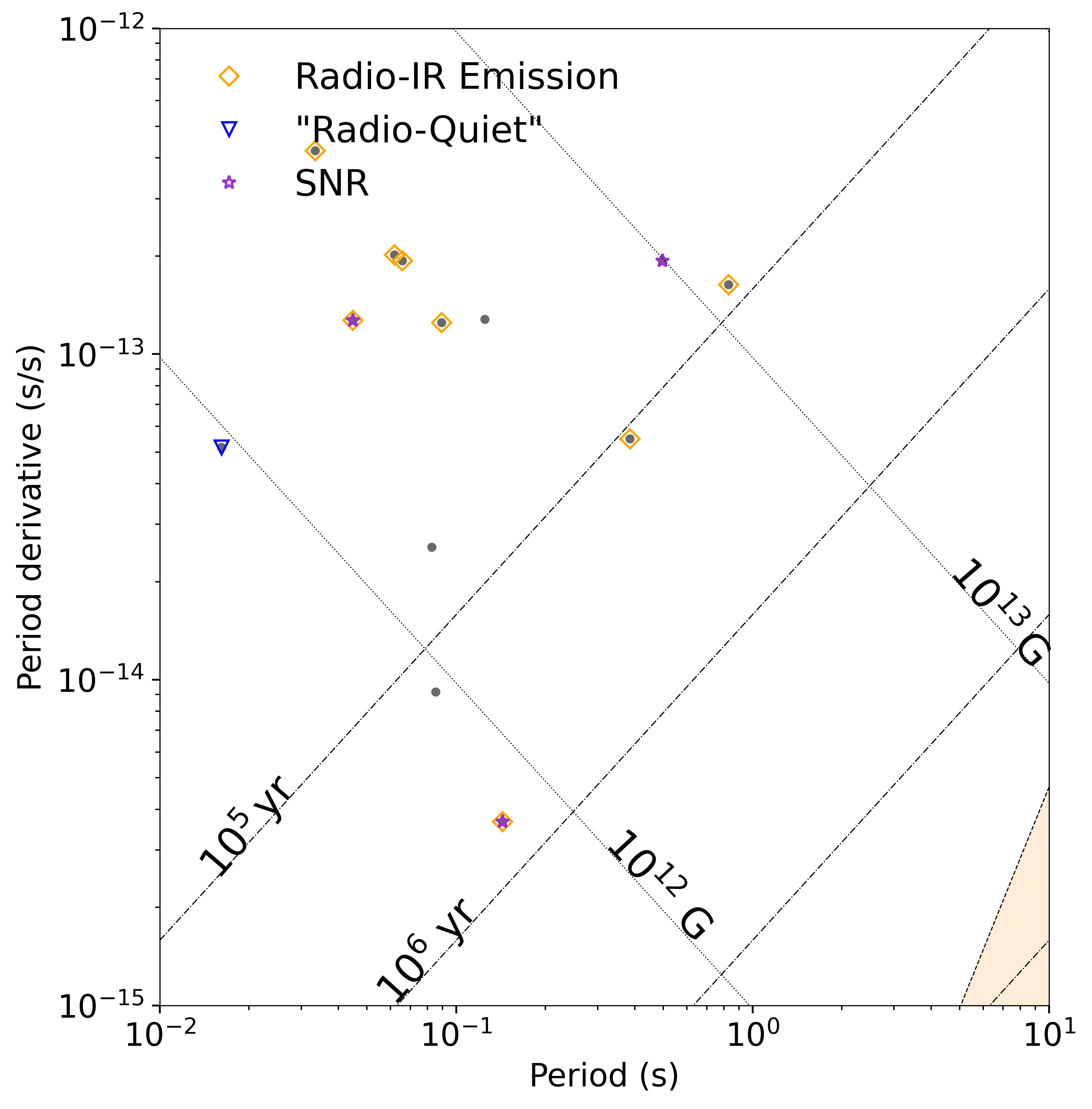}
    \caption{$P\dot{P}$ diagram for the population of pulsars studied in this work. The dashed and dotted lines correspond to characteristic ages and spin-down inferred magnetic fields respectively \citep{2018JOSS....3..538P}. The orange region in the bottom right is the part of the $P\dot{P}$ diagram beyond the death line, where pair creation and cascades are no longer efficient, thus pulsars stop emitting in radio \citep{1993ApJ...402..264C}.}
    \label{fig:2}
\end{figure}

In this section we discuss the sample of neutron stars on which we focus our analysis. We chose a sample of young neutron stars, since we expect them to have small spin periods and strong magnetic fields, that are associated with SNRs. We also needed an indication for their current space velocity, so we focused on collecting their radial or transverse velocity. Since the determination of the radial velocity is often difficult due to the intense broadening of the spectral lines because of the strong magnetic field of a pulsar \citep{lovis2010radial}, we focus on the transverse velocity ($V_t$) which we derive from the ATNF catalogue \citep{2005AJ....129.1993M}, along with other parameters such as the magnetic field strength, the current spin period of each pulsar and the periods derivative. Since these pulsars are associated with SNRs it was important to have an estimate of the age of each SNR ($\tau_{SNR}$). For this analysis we used a series of previous works \citep{igoshev2022initial, Suzuki_2021, 2012AdSpR..49.1313F} and the SIMBAD astronomical database \citep{wenger2000simbad}. We have selected the sources in this sample under the following criteria: either the transverse velocity along with the age of the hosting supernova remnant is known, or the braking index has been measured using a timing solution. We focus on the value of the braking indices, since  it is a good indication for the different spin-down mechanisms that can decelerate a pulsar. This quantity constitutes a guide to draw conclusions
on the evolution of the rotation of a neutron star  \citep{livingstone2006braking}.  In section 2.5 this characteristic quantity is discussed further. The two samples are presented in tables \ref{3.2t}, \ref{3.1t}. In Figure \ref{fig:2} we show the $P\dot{P}$ diagram the sample \citep{2018JOSS....3..538P}, which consists of 13 pulsars. 

In this diagram we can see that according to the ATNF catalogue, the sample includes pulsars that are associated with SNRs except two of them, which appear as black dots. These two neutron stars are J0205+6449 and J1809-1917, which according to the SNRcat \citep{2012AdSpR..49.1313F} are associated to SNRs as shown in Table \ref{3.2t}. In addition two of these stars are characterized as "Radio Quiet" (appear as blue triangles) i.e., we do not observe them through the radio emission, while 7 of them have a strong emission in radio waves and are referred to as "Radio-IR Emission” (shown as yellow diamonds). Most of them are very young pulsars, with magnetic fields that reach up to $10^{12}$ $\rm{G}$. 

\begin{table}\label{table:1}
\centering
\setlength{\tabcolsep}{2.2pt}
\renewcommand{\arraystretch}{2.4}
\hspace{2cm}
\caption{The sample population of 13 neutron stars associated with supernova remnants.} 
\begin{tabular}{c c c c c c  } 
\hline 
\thead{PSR\\NAME} & \thead{SNR\\}  & \thead{$P$\\$\rm{(ms)}$} &  \thead{$\dot{P}$\\$(\rm{s/s}) (10^{-13})$} & \thead{$B$\\$(\rm{G})(10^{12})$} & \thead{$\rm{V_t}$\\($\rm{km/s}$)} \\
[0.5ex] 
\hline 
J0205+6449 & G130.7+03.1 & 66 & 1.94 & 3.61 & 22.7  \\ 
J0534+2200 & Crab & 33 & 4.21 & 3.79  & 142.67  \\ 
J0537-6910 & N157B & 16 & 0.51 & 0.92 & 634 \\ 
J0538+2817 & G180.0-01.7 & 143 & 0.03 & 0.73 & 356.7 \\
J0659+1414 & Monogem Ring & 385 & 0.54 & 4.65 & 60.3 \\
J0835-4510 & Vela & 89 & 1.25 & 3.38 & 76.97 \\ 
J1731-4744 & G343.0-6.0 & 830 & 1.64 & 11 & 500.59 \\ 
J1801-2451 & G5.4-1.2 & 125 & 1.28 & 4.04 & 199.15 \\ 
J1809-1917 & G005.4-01.2 & 83 & 0.25 & 1.47 & 828 \\ 
J1813-1749 & G012.8-00.0 & 45 & 1.27 & 2.41 & 411.3 \\ 
J1833-0827 & G023.5+00.1 & 85 & 0.09 & 0.89 & 715.58 \\ 
J1833-1034 & G21.5-0.9 & 62 & 2.02 & 3.58 & 125 \\ 
J2337+6151 & G114.3+0.3 & 495 & 1.93 & 9.91 & 164 \\ 
\hline 
\label{3.2t}
\end{tabular}
\end{table}

\subsection{Initial spin period of pulsars}
In this section we discuss the initial spin period of pulsars since it is a key parameter in our analysis. Both for the rocket model and for our alternate model, the smaller the initial spin period, the greater will be the kick velocity, as it is during this stage of a neutron star's life that the electromagnetic acceleration will be the highest, due to rapid rotation. Specifically, \citet{2001ApJ...549.1111L} in their analysis, used an initial spin period of 1$\rm{ms}$ as \citet{xu2022birth} to test the electromagnetic kick. 
In addition, \citet{haensel1999minimum} made an estimation of the minimum spin period ($P=0.288$ $\rm{ms}$) without however taking into account a specific equation of state for the neutron star. From population studies, \citet{igoshev2022initial} found that for neutron stars in SNRs, the initial spin period can be described by a log-normal distribution with $\mu_{log(P/s)}=-1.04$ i.e. $P= 0.091$ $\rm{s}$ and $\sigma_{log P}=0.53$. 

Contrary to these studies, by using radio polarization, \citet{lorimer1993pulsar} proposed that radio pulsars are born with spin periods of 0.02-0.5 s, which indicate a slow rotation, which would not allow for the rocket effect to have any impact. In general, the determination of the initial spin period of neutron stars is a rather difficult, even for pulsars that are associated with SNRs for which at least an independent estimate of the age exists. It is known that the initial rotational period should be fast for the young pulsars, and especially magnetars, to have such strong magnetic fields, for which there is strong support now \citep{raynaud2020magnetar}.

As mentioned above, \citet{igoshev2022initial} constrained the distribution for the initial spin periods for a population of pulsars associated with SNRs. They found that for the pulsars whose characteristic age is higher than the SNR age they follow a distribution with mean value of $0.091$ s, whereas for several sources, the SNR age is longer than the characteristic age and such an estimate is not feasible. For our analysis we use the same methodology as \citet{igoshev2022initial}: we calculate the initial spin period we use equation \ref{tau_c} but instead of the characteristic age of the pulsar, we replace it with the age of the SNR.
For the cases where characteristic age is greater than the age of the SNR and the equation above would lead to a negative square root, we assume that the initial rotation period is as low as $P_{in} =2$ $\rm{ms}$ or $P_{in}=1$ $\rm{ms}$. The fact that the square root is negative could be explained by two scenarios: the SNR age is overestimated or the pulsar was slowing down with a smaller rate in the past. Based on that we calculate the kick velocity derived from equation \ref{10} using the approach described above and the methodology described in the previous paragraph using equation \ref{eq1}.

\subsection{Braking indices}

As mentioned above one significant parameter is the braking index for a pulsar. Specifically we can, depending on its value, draw some conclusions about the mechanism that leads to the spinning-down of neutron stars. Therefore in this paragraph, we will focus on the braking index within the framework of the theory of electromagnetic kick.  Considering that the angular velocity $\omega$ of neutron stars evolves as ${\dot{\omega}}\approx  \omega^n$, where $n$ is the braking index, then we can derive to the equation which gives the braking index \citep{livingstone2006braking}:
\begin{equation*}
    n = \frac{\omega {\ddot{\omega}}}{{\dot{\omega}}^2},
\end{equation*}
where $\ddot{\omega}$ is the second derivative of the angular velocity. Assuming that pulsar spin-down is due to a dipolar magnetic field, while other physical parameters (moment of inertia, magnetic field strength, angle between the magnetic dipole axis and the spin axis) remain unchanged, leads to the canonical value $n=3$. However, as shown in Table \ref{3.1t} several pulsars have braking indices that deviate from the canonical value with extreme examples being pulsar J1809-1917 with a braking index $23.5$ and pulsar J1801-2451, with a braking index $1.1$ \citep{2020}. In reality, the magnetic field of a pulsar may have a complex structure, which can result to a different value for the braking index. Also the spin-down rate may be affected by other factors, such as the presence of a relativistic wind \citep{menou2001stability}, moment of inertia evolution \citep{ho2012rotational}, magnetic field evolution \citep{gourgouliatos2015hall} or emission of gravitational waves \citep{2016EPJC...76..481D}. These additional factors can cause the pulsar to spin down at a rate that leads to a braking index $n\neq 3$. Here, we explore whether an offset magnetic field can be related to the measured braking indices. 

For this analysis we consider another sample of neutron stars for which the braking index values are known, to compare these values with the ones that arise if pulsars have a displaced magnetic field from their center. We use as a guide the work of \citet{10.1093/mnras/stw2050}, who developed analytical solutions for the electromagnetic fields outside a neutron star for a vacuum magnetosphere and an off-axis dipolar or multipolar magnetic field. The methodology used there is similar to the one we follow here, since first a multipole expansions for the fields is performed, with the use of spherical harmonic functions, then integrating for the Poynting flow the spin-down luminosity is found and eventually the force exerted on the star due to the emission of radiation. Assuming that the deceleration of the star is solely due to the emission of electromagnetic radiation due to the existence of a magnetic dipole, we know that the following relationship will apply, 
\begin{equation}
    L_{sd}=-I\omega{\dot{\omega}},
\end{equation}
where $L_{sd}$ is the spin-down luminosity and $I$ is the moment of inertia of the star. Deriving the equation above in terms of time, 
\begin{equation}
   {\dot{L_{sd}}} =-I{\dot{\omega}^2}-I \omega {\ddot{\omega}} \Rightarrow  \frac{{\dot{L_{sd}}}}{{\dot{\omega}^2}}= I-In 
   \Rightarrow \dot{L_{sd}}= \frac{\omega {\ddot{\omega}} L_{sd}}{n \omega {\dot{\omega}}}+ \frac{\omega {\ddot{\omega}} L_{sd}}{ \omega {\dot{\omega}}},
\end{equation}
which finally gives us the braking index as a function of the spin-down luminosity:
\begin{equation}
    n=\frac{dL_{sd}}{d\omega} \frac{\omega}{L_{sd}}-1.
\end{equation}

We can rewrite the above expression as:
\begin{equation}
    n=\frac{dL_{sd}}{dh} \frac{h}{L_{sd}}-1,
\end{equation}
where $h=R/R_{LC}$ and the relationship $R_{LC}=c/\omega$ was used, $R_{LC}$ is the radius of the light cylinder and $R$ is the stellar radius. 

According to \cite{10.1093/mnras/stw2050}, the final expressions for the braking index: 
\begin{equation}\label{n1}
     n(\alpha=0^o)=5+ \frac{1}{7} h^2 \epsilon^2(97-95\cos{2\delta}),
 \end{equation}
 \vspace{-0.5cm}
\begin{equation}\label{n2}
     n(\alpha=90^o)= 3+ \frac{1}{5} h^2 \epsilon^2(93+\cos{2\beta}) \sin^2\delta. 
 \end{equation}

Here $\alpha, \beta, \delta$ angles are shown in Figure 1 in \citet{10.1093/mnras/stw2050}. Specifically, the angle $\alpha$ is the obliquity of the magnetic axis, so for $\alpha=0^o$ we refer to an aligned rotator, while for $\alpha=90^o$ we consider an orthogonal rotator. Moreover the angle $\beta$, is the angle out of the meridional plane and $\delta$ is the angle between the rotation axis and the line joining the dipole to the centre.What we are interested in, is to compare the value of the braking index obtained from the above relations with the measured values we have so far for a series of pulsars located in supernova remnants. We focus on 16 neutron stars to investigate whether the rocket effect can be applied to them. All pulsars in this sample belong to supernova remnants and have short rotation periods and young ages.

We aim to find the extreme values of the braking index resulting from these relationships, meaning the maximum and minimum values. First of all the coefficient $\epsilon=s/R$ will vary between values [0,1], since the distance $s$ cannot be bigger than the stars radius. Accordingly, the term ${h=R/R_{lc}}$ will clearly always be less than unity, but it will also depend on the period of each star since the radius of the light cylinder ${R_{lc}=cP/2\pi}$. Therefore we find the maximum and minimum value of $h$, for the pulsar with the shortest and longest rotation period respectively, assuming that the stellar radius equals to 12 $\rm{km}$. The shortest rotation period for the sample population of stars is $P=16$ $\rm{ms}$ for J0537-6910 pulsar for which we calculate that the maximum value of $h$ is ${h_{max}=0.01}$, while the longest one is $P= 814$ $\rm{ms}$ for the pulsar J1632-4818, for which the $h$ minimum value is $\rm{h_{min}=0.26 \times 10^{-3}}$.

\begin{table}
\caption{Braking index of the 16 pulsars of our catalogue along with the corresponding references.} 
\setlength{\tabcolsep}{3pt}
\renewcommand{\arraystretch}{2.2}
\centering 
\begin{tabular}{c c c c} 
\hline 
Pulsar Name & n & Article  \\ [0.5ex] 
\hline 
J0534+2200 (Crab) & 2.51 & \cite{10.1093/mnras/233.3.667}\\
J0537-6910 & 7.4 & \cite{Ferdman2017TheGA}\\
J0540-6919 & 2.12 & \citet{Wang_2020}\\
J0659+1414 & 14.43 & \citet{https://doi.org/10.1046/j.1365-8711.1999.02737.x}\\
J0835-4510 (Vela) & 2.81  & \cite{2020}\\
J1119-6127 & 2.65 & \cite{2010}\\
J1513-5908 & 2.82 & \cite{2020}\\
J1632-4818 & 6 & \cite{2020}\\
J1640-4631 & 3.15 & \cite{refId01}\\
J1731-4744 & 2.7 & \cite{refId0}\\
J1801-2451 & 1.1 & \cite{20161}\\
J1809-1917 & 23.5 & \cite{2020}\\
J1833-0827 & 15 & \cite{2020}\\
J1833-1034 & 1.8 & \cite{10.1111/j.1365-2966.2012.21380.x}\\
J1846-0258 & 2.65 & \citet{ho2012rotational}\\
J2337+6151 & 8.60 & \citet{https://doi.org/10.1046/j.1365-8711.1999.02737.x} \\ [1ex] 
\hline 
\label{3.1t}
\end{tabular}
\end{table}

Next, by taking into account the maximum and minimum values of the trigonometric functions that appear in equations \ref{n1}, \ref{n2}, we find that the braking index will always be very close to 3 or 5. Specifically for the neutron star J0537-6910, if we assume an inclination angle of ${\alpha=0^o}$ the maximum value of the braking index, that corresponds to $\epsilon=1$, is $\rm{n_{max}=5.000023}$ while the minimum value will be $\rm{n_{min}=5}$.
 For the case where ${\alpha=90^o}$ we will have $\rm{n_{max}=3.00018}$ and $\rm{n_{min}=3}$. 
As the measured braking indices deviate significantly from the canonical values, 
even if the rocket effect operated in such systems its impact would not be significant, affecting the fourth decimal point, or even less. Thus, the measured values seem to be caused by effects beyond the framework of the rocket effect model.

The above result may be due to the fact that
the previous study assumes that a neutron star is emitting only electromagnetic dipole radiation which is the only cause of neutron star deceleration. If the emission of gravitational waves is included in this analysis, the above expressions for the braking index will be different, and maybe give more realistic results. Additionally, the assumption that the pulsar has a non rotating vacuum magnetosphere, while it makes the analysis more manageable, is a simplification from the natural state of the magnetosphere, and can affect the braking indices.

\section{Results}\label{3}
\subsection{Kick Velocities}

 Here we present our results for the kick velocity obtained by the rocket effect model. Our results are classified according to the model that has been used and the value of the initial spin period. More specifically, our models are named as $V_{kick}^{HT}$ where the kick velocity is derived by using the model of \citet{Harrison1975AccelerationOP}, and $V_{kick}^{AL}$ from the values exerted by the time-dependent model shown in \ref{sec:alternative}, respectively. The initial spin period calculated by equation \ref{tau_c}, following the analysis of \citet{igoshev2022initial} and are shown in Table \ref{3.3}. For our calculations, we assume that the radius of the neutron star is 12 $\rm{km}$ and that its mass is 1.44 $M_{\odot}$. Moreover we set the distance of the magnetic axis from the spin axis $s=10$ $\rm{km}$, and we consider all these parameters to be constant over time, as well as the magnetic field of the star.

\begin{table}
\centering
\setlength{\tabcolsep}{4.5 pt}
\renewcommand{\arraystretch}{2.4}
\caption{The results of the kick velocity for our population of 13 pulsars.The velocities are in $\rm{km/s}$, and they have been measured by both models.} 
\begin{tabular}{c c c c c c}
\hline
        Pulsar Name & \thead{$\rm{P_{in}}$\\$\rm{(ms)}$} & \thead{$\rm\tau_{snr}$\\$(\rm{kyr})$} & \thead{$\rm{V_{kick}^{HT}}$\\$(\rm{km/s})$} & \thead{$\rm{V_{kick}^{AL}}$\\$(\rm{km/s})$}
        \\ \hline
        J0205+6449 & 39 & 7 & 0.01 & 0.01  \\ 
        J0534+2200 & 16 & 0.96 & 0.30 & 0.16  \\ 
        J0537-6910 & 35 & 4.93 & 179.65 & 101.77 \\ 
        J0538+2817 & 2 & 34.80 & 179.99 & 123.41  \\ 
        J0659+1414 & 2 & 128 & 179.99 & 101.66 \\ 
        J0835-4510 & 2 & 1.90 & 179.99 & 101.69  \\ 
        J1731-4744 & 2 & 20 & 179.99 & 101.65 \\ 
        J1801-2451 & 2 & 40 & 179.99 & 101.37 \\ 
        J1809-1917 & 2 & 1.90 & 179.99 & 101.35 \\ 
        J1813-1749 & 40 & 1.20 & 0 & 0 \\ 
        J1833-0827 & 29 & 130 & 0.05 & 0.03 \\ 
        J1833-1034 & 50 & 1.67 & 0 & 0 \\ 
        J2337+6151 & 2 & 7.7 & 179.99 & 102.62\\ \hline
        \label{3.3}
\end{tabular}
\end{table}

In Table \ref{3.3}, the results for the kick velocity are quite different for each case. If we compute the initial spin period from equation \ref{tau_c} the model of HT seems to give bigger kick velocities than our alternate model, although for some pulsars the kick is almost zero, like PSR J1813-1749 and J1833-1034. For these pulsars the electromagnetic rocket effect does not seem to have an application, even though both pulsars have quite large transverse space velocities. We should take into consideration the strength of the magnetic field or the spin period of these neutron stars. For example for the pulsar J1813-1749, even though we see a large transverse velocity 411 $\rm{km/s}$, it seems that this pulsar is not a millisecond pulsar, and the integration takes place for a small timescale, comparatively to the other pulsars, since the age of the associated SNR is 1200 years. In general, the larger the age of the SNR, the bigger values we predict for the kick velocities. Now, assuming that the initial spin period is 1 $\rm{ms}$, the kick velocities from our alternate model seem to change, since the mean kick velocity for the sample now is 731 $\rm{km/s}$. For the HT model, given initial spin periods of 1 $\rm{ms}$, the kick velocity is respectively 2000 $\rm{km/s}$.

As a next step we pursued, to see which is the most likely value of distance $s$, so the kick velocity expected by the rocket effect to be equal to the transverse velocity of each pulsar (see Table \ref{3.2t}). We solve equation \ref{10} for $s$ and assume that the initial spin period is 1 $\rm{ms}$ for all pulsars.
\begin{table}
\centering
\setlength{\tabcolsep}{4.5 pt}
\renewcommand{\arraystretch}{2.4}
\caption{In this table is shown the value of the displacement $s$ of the axis of the magnetic field and the center of the star, so that the star has a kick velocity equal to the transverse component of the current space velocity $\rm{V_t}$. For these calculations we take into consideration an initial spin period of 1 $\rm{ms}$ for all pulsars.} 
\begin{tabular}{c c c}
\hline 
        Pulsar Name & \thead{Vt\\($\rm{km/s}$)} & \thead{s\\$\rm{(km)}$} \\ \hline
        J0205+6449 & 22.7 & 0.24 \\ 
        J0534+2200 & 142.67 & 1.46 \\ 
        J0537-6910 & 634 & 8.25 \\ 
        J0538+2817 & 356.7 & 4.86 \\ 
        J0659+1414 & 60.3 & 0.82 \\ 
        J0835-4510 & 76.97 & 1.04 \\ 
        J1731-4744 & 500.59 & 6.83 \\ 
        J1801-2451 & 199.15 & 2.72 \\ 
        J1809-1917 & 828 & 9.77 \\ 
        J1813-1749 & 411.3 & 1.84 \\
        J1833-0827 & 715.58 & 8.43 \\ 
        J1833-1034 & 125 & 0.73 \\ 
        J2337+6151 & 164 & 2.23 \\ \hline
        \label{3.5t}
    \end{tabular}
\end{table}
As seen in Table \ref{3.5t}, we find acceptable values for the parameter $s$, since all results are less than the allowable limit that is the radius of the star (12 $\rm{km}$). The smallest value for displacement appears to correspond to the star J0205+6449 and is $s=0.24$ $\rm{km}$, while the largest displacement value is 9.77 $\rm{km}$ for the star J1809-1917, which has the highest  proper motion velocity of all the population of neutron stars. Although a displacement of such large scale does not seem realistic, it seems to be the most extreme value in our calculations. 

To investigate whether the rocket effect needs to operate for a significant amount of time to provide the observed velocity, we have integrated equation \ref{v_AL_time} for shorter time intervals. We found that within the first 10 years of a pulsar's life, the rocket effect accelerates the star to at least $50\%$ of the required  velocity. So, most of the acceleration should take place in the early part of a neutron star's life.

\subsection{Markov Chain Monte Carlo Analysis}

Next, we perform a MCMC statistical analysis, to get the posterior probability densities for the distance $s$ and the initial spin period $P_{in}$, so that the kick velocity of the rocket effect mechanism is equal to the current transverse velocity of each pulsar. We make the assumption that the current spatial velocity of this population, is solely due to the rocket effect, and we do not take into consideration other kick mechanisms. Moreover we neglect the gravitational wave emission for our population. For this statistical analysis we use the emcee package \citep{2013PASP..125..306F}, which is a Python implementation of the affine-invariant ensemble sampler for MCMC. Here multiple walkers run in parallel but allowed to interact in such a way that they can adapt their proposal densities.

The emcee package is an algorithm that performs equally well under all linear transformations, so it is insensitive to covariances among parameters. The free parameters for this analysis are the initial spin period of pulsars and the distance of the magnetic axis from the star's center ({$P_{in},~ s $}). These parameters are given by random values in our parametric space, since we only give them the initial values so that the algorithm begins. More specifically we assume that the initial spin period and the distance are described by an uniform distribution. For the former, the accepted values are in the range of [0,0.5] s, since we believe that this mechanism appears only in pulsars with fast rotation,  while for the latter the range is [0, 10] $\rm{km}$. Moreover we use an optimizer to minimize the initial values that are needed for the MCMC analysis. This algorithm will finally give us the posterior probabilities of these parameters so that the kick velocity for the population due to the rocket effect agrees with the values of the transverse component of the current space velocity.

\begin{figure}
    \includegraphics[width=1\linewidth]{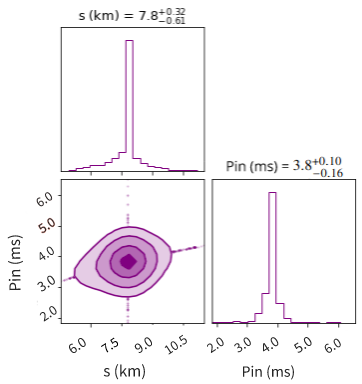}
    \caption{The corner plot derived after our MCMC analysis, for our sample population. The 2D posterior probability density distributions for our parameters are shown in lighter and darker purple regions while the 1D posterior probability distributions are shown on the diagonal. On the top of the diagonal plots shows the mean estimated value and the standard deviation $\sigma$.}
    \label{fig:4}
\end{figure}

As shown in corner plot \ref{fig:4}, the MCMC analysis gives the 2D posterior probability density distributions for our parameters $s, ~P_{in}$. To produce this diagram we assume that the neutron stars of our sample population have a radius of 12 $\rm{km}$ and a mass of 1.44 $M_\odot$. The rest parameters are kept stable for each pulsar, such as the magnetic field, the age of the SNR and the first derivative of the current spin period (See Tables \ref{3.2t},\ref{3.3} ). Since we know the values for these parameters with great accuracy, we do not set them as free parameters.

Finally, the emcee package, returns the posterior probabilities. As shown in Figure \ref{fig:4}, the median values and their errors for our parameters appear on the top of the 1D distributions. For the distance $s$ the mean value is $7.8^{+ 0.32}_{-0.61}$ $\rm{km}$ and for the initial spin period $ 3.8 ^{+ 0.10}_{-0.16}$ $\rm{ms}$. It seems that the distance $s$ is quite large, but this value is about a whole population of pulsars with different properties. Comparing to the values of Table \ref{3.5t}, it seems that a large distance could be acceptable, since for some pulsars, such as J1833-1034, J0537-6910 and J1809-1917, the distance $s$ is larger than 8 $\rm{km}$. Moreover, the results from the MCMC analysis, are realistic, within the acceptable range, and agree with the assumption that the rocket effect mechanism applies to fast rotating neutron stars, since our values for $P_{in}$ are very small.

\subsection{Case studies}\label{4f}
{\subsubsection{{The pulsar J0030+0451}}}

One remarkable case is the millisecond isolated pulsar J0030+0451, for which a detailed mapping of the thermally emitting regions was feasible. This study was performed independently by two  research groups \cite{miller2019psr,riley2019nicer}, and was aimed to determine the mass and radius of the star, in addition to mapping the thermal regions. For this studies both groups relied on  observational data from the NICER X-ray (Neutron Star Interior Composition Explorer) telescope and reported
strong evidence for the existence of a multipolar magnetic field for this pulsar, since the predicted geometry of the hot spots do not indicate a centred dipole field. 
Later, based on these results, \citet{Kalapotharakos_2021} estimated the structure for the magnetic field of this pulsar, so that their simulated light curves for this pulsars in gamma and X-rays, would agree with the ones that NICER had provided, justifying the geometry of the hot spots. In this study, it was confirmed that the magnetic axis should be off-centred and that the magnetic field consists of dipole and quadrupole components. This significant result, confirms the theory of multipolar stellar magnetic fields, and gives as a reason to test the rocket effect mechanism on the J0030+0451 pulsar. Even though, this neutron star is not associated with an SNR, so it is not a young magnetar, and has a small transverse velocity, it could still be a good candidate to test the rocket effect mechanism, since its offset magnetic dipole field and the fact that it has a very small spin period, of 4.87 $\rm{ms}$ (See Table \ref{4.2t}).

In their study \citet{Kalapotharakos_2021}, found different offset morphologies for the magnetic field, and many families of solutions for static vacuum field models and force free models. Specifically they found that for some of these models the dipole and quadrupole moment are slightly offset from the star's center. For example, starting with a magnetic field which consists of the sum of an offset dipole and quadrupole moment, for the case of a force free field, the coordinates for the dipole magnetic axis are $(x, y, z)=2.08, -2.73, -4.94$ $\rm{km}$. So it seems that the displacement of the magnetic field is small from the stars center, but is enough for the rocket effect to take place. 

To test if the rocket effect mechanism could at some point, have an effect on this pulsar, we calculated the kick velocity of J0030+0451, using equation \ref{10}. For this calculation, we considered that the initial spin period is 2 $\rm{ms}$, the distance $s$ of the magnetic axis is 2.08 $\rm{km}$, based on \citet{Kalapotharakos_2021}, and we integrated for a period of time, equal to the characteristic age of the pulsar, which is ${\tau_c} = {7.58 \times 10^9 \rm{yrs}}$ . Finally, we estimate that the kick velocity should be 228 $\rm{km/s}$, if the rocket effect had acted upon the star. This velocity is much bigger than the current true transverse velocity of J0030+0451, which does not exceed 17 $\rm{km/s}$. For this pulsar we did an MCMC analysis, as we did for the previous population of pulsars, and by using the values that appear in Table \ref{4.2t}, we found that the distance $s$ should be 3.02 $\rm{km}$, and the initial spin period 3.52 $\rm{ms}$, for the kick velocity to be equal to the current transverse velocity of this pulsar.

Therefore, it is very likely that the rocket effect mechanism has no application in this pulsar. Even though it is a very fast rotating pulsar, and it seems that its magnetic field has offset components, we would expect that if a kick mechanism  had been applied to this neutron star, its space velocity would be bigger. We believe that at some point a kick mechanism was applied on this pulsar, but the results of this acceleration cannot be seen anymore. It is very likely, that this isolated pulsar, had been through a spun up by a companion at its first years, via accretion of matter. This accretion could lead to the increase of the pulsars angular momentum, finally forming a millisecond pulsar. Moreover this accretion of mass could be one of the reasons that led to this multipolar components of the stellar magnetic field, that we see through the NICER observations. Assuming that the J0030+0451 pulsar had a similar magnetic field structure as its progenitor, which could be described by a dipole filed, it is possible that a violent accretion along with a fast spun up, could have some effect on the form of the magnetic field.  As proposed by \citet{1989ApJ...346..847C}, young pulsars go through  violent accretions of mass coming from a companion star, which could bury the stellar magnetic field in the crust. So the fact that this pulsar is no longer a magnetar could be explained by a buried magnetic field, because of the destruction of a near star or planet. Thus, the fact that this millisecond has most likely undergone a spin-up phase, makes the characteristic age irrelevant for this calculation. 
\begin{figure}
    \includegraphics[width=1\linewidth]{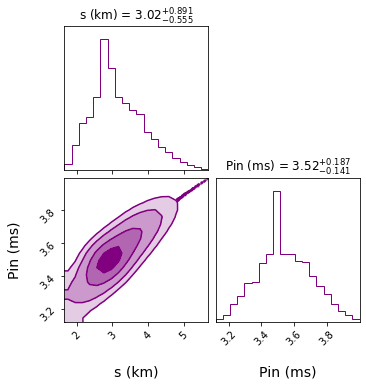}
    \caption{The corner plot derived after our MCMC analysis, this time for the pulsar J0030+0451. The 2D posterior probability density distributions for our parameters are shown in lighter and darker purple regions while the 1D posterior probability distributions are shown on the diagonal. On the top of the diagonal plots shows the mean estimated value and the standard deviation $\sigma$.}
    \label{fig:j00}
\end{figure}

It should be mentioned that this model, as well as the initial model of \citet{Harrison1975AccelerationOP}, is for an offset dipole magnetic field, and does not take into consideration higher multipole components. The J0030+0451 is believed to have quadrupole and higher components of the magnetic field, so the expressions for the kick velocity should be adjust to these conditions, since the luminosity of the pulsar could not be described by equation \ref{4}. So it seems, that in general the rocket effect mechanism, should be redefined to fit cases of neutron stars, which are likely to have multipolar magnetic field components.

\subsubsection{The pulsar J0538+2817}

The pulsar J0538+2817, which is associated with the SNR G180.0-01.7 (S147), is a unique case of neutron star. First of all, for this pulsar, the three-dimensional spin- kick alignment was recently confirmed for the first time. In their analysis, \citep{2021NatAs...5..788Y} used a scintillation
method to estimate the radial velocity of this pulsar by taking into consideration observations from the Five-hundred-meter Aperture Spherical Radio
Telescope (FAST). They measured the inclination angle between the pulsar velocity and the line of sight as $\zeta_v = 110^{\circ}$, while by adopting a polarization method, they found the inclination angle of the spin axis and the line of sight as $\zeta_{pol} = 118^{\circ}$. 

In general, this is a phenomenon that has been proposed as a result of either the electromagnetic kick mechanism \citep{{10.1111/j.1365-2966.2005.09669.x}} or another kick mechanism  \citep{janka2022supernova} such as the neutrino-driven kick mechanism, and gives the maximal values for the kick velocity. Moreover, \citet{xu2022birth} suggested that this alignment could be due to the rocket effect mechanism for this neutron star, and calculated the expected kick, based on the model of \cite{Harrison1975AccelerationOP}. Considering an initial spin period of 1ms and a distance $s$ of 7 $\rm{km}$ they found that the kick velocity is 400 $\rm{km/s}$, which is very close to the current velocity of this pulsar (see Table \ref{4.2t} ). On the other hand, \citet{janka2022supernova}, proposed that the observed properties of J0538+2817 pulsar are not in favor of the rocket effect, since it is more likely that its initial spin period is bigger than 1 $\rm{ms}$.
Here, we suggest that this pulsar seems to be a good candidate to test the rocket effect model, since it has a small spin period, and a transverse velocity high enough, to assume that a kick mechanism  have probably acted upon the pulsar at some point in its life. Nonetheless, the magnetic field of J0538+2817, is not that strong while we expect the rocket effect to have a better application in magnetars. But it is possible that this pulsar was a magnetar at its early life, and have experienced a magnetic field decay as time passed, due to accretion of mass. As mentioned above, \citet{1989ApJ...346..847C} argued that  accretions of mass coming from a companion star, buries the stellar magnetic field in the crust. So it is possible that the electromagnetic kick mechanism had acted on the pulsar while it was newborn and still had a stronger magnetic field.

\begin{table}
\centering
\setlength{\tabcolsep}{2.5pt}
\renewcommand{\arraystretch}{2.4}
\caption{Characteristic quantities of the pulsars J0538+2817 and J0030+0451 that are used in the MCMC analysis.} 

\begin{tabular}{c c c} 
    \hline
        Parameter & J0538+2817 & J0030+0451\\ 
        \hline
        Period ($\rm{ms}$) & 143 & 4.87 \\ 
        Period Derivative ($\rm{s/s}$) & $\rm{3.6\times10^{-15 }}$ &  $\rm{1.02\times10^{-20}}$ \\
        $\rm{\tau_c}$ ($\rm{yrs}$) & $\rm{6.18\times10^5 }$ &  $\rm{7.58 \times 10^9}$ \\ 
        $\pi$ ($\rm{mas}$) & 0.72 & 3.3 \\ 
        $\mu$ ($\rm{mas/yr}$) & 60 & 58\\ 
        $V_t$ ($\rm{km/s}$) & 357 & 8-17 \\ 
        $D$ ($\rm{kpc}$) & 1.2 & 0.32 \\
        $B$ ($\rm{G}$) & $\rm{7\times10^{11}}$ & $\rm{2x10^8}$ \\ 
        Mass & $\rm{1.44 M\odot}$ & -\\ 
        $R$ (km) & 13.02 & -\\ \hline
        \label{4.2t}
    \end{tabular}
\end{table}

\begin{figure}
    \includegraphics[width=1\linewidth]{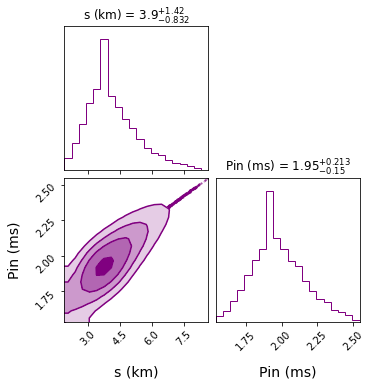}
    \caption{The corner plot derived after our MCMC analysis, this time for the pulsar J0538+2817. The 2D posterior probability density distributions for our parameters are shown in lighter and darker purple regions while the 1D posterior probability distributions are shown on the diagonal. On the top of the diagonal plots shows the mean estimated value and the standard deviation $\sigma$.}
    \label{fig:6}
\end{figure}

Even though this pulsar is in our catalogue depicted in Table \ref{3.3}, we wanted to make a separate MCMC analysis for this case. As for the analysis of our sample population, we considered an uniform distribution for the distance $s$ and the initial spin period, with values that are in the same range as our previous analysis. Our results can be seen in Figure \ref{fig:6}. In order to have similar values for the current transverse velocity of the pulsar and the kick that is expected from the rocket effect mechanism, we estimate that the distance $s$ should be $3.90^{+ 1.42}_{-0.83}$ $\rm{km}$ and the initial spin period $ 1.95 ^{+ 0.21}_{-0.15}$ $\rm{ms}$. Our result for the distance $s$ of the magnetic axis, has a declination from the result in Table \ref{3.5t}, where we found that the distance $s$ should be 4.86 $\rm{km}$, from the star's center. This declination is completely due the value of the initial spin period of the pulsar, since for the results in table \ref{3.5t} we consider a strict value of 1 $\rm{ms}$. 

According to our MCMC analysis, it seems that the initial spin period should be very small, so that the rocket model has an impact on the neutron star. It should be mentioned that, \citet{2003ApJ...593L..31K} presented a timing proper-motion measurement for this pulsar and estimated that its initial period should be 139 $\rm{ms}$, a value which comes with many constrains on the kick mechanisms that could be applied to this pulsar \citet{Romani_2003}, since it describes a very slow rotating pulsar. We argue that this pulsar should have a very small initial period, so that its large transverse velocity can be justified. It is very likely that, after the spun up of the pulsar, due to its electromagnetic and gravitational emission, follows a spin down. In their work, \citet{xu2022birth} proposed that the spin period of the pulsar evolves proportionally with time, so we expect the pulsar to slow down eventually.

\section{Discussion and Conclusions}

\label{Discussion}

In this work, we have explored the impact of the rocket effect to observed properties of neutron stars. We have focused on three main aspects: the relation of the rocket effect to pulsar braking indices, its effect to the observed spatial velocities of pulsars and its application to two specific cases for which we have strong indications that an off-centre dipole magnetic field or even multipolar field is present. 

An important issue is to understand what  are the conditions that lead to offset and multipolar magnetic field configurations. For example \cite{ruderman1991neutron} proposed that neutron
superfluid vortex lines that travel towards the star's crust, can modify the magnetic field configuration depending on the strength and the spin down of the pulsar. Moreover in order to explain the magnetic field structures at the surface of
neutron stars, many studies have been focused on the Hall drift effect, which can be associated with a magnetic field that has multipole components \citep{gourgouliatos2018modelling, geppert2014creation}. We note that axisymmetric models cannot have offset fields, however, in 3-D studies this is possible. Even if the exact details of the mechanism are yet to be understood, the existence of multipolar and offset dipolar fields in J0030+0451, makes this scenario likely.

Regarding the impact of the rocket effect to braking indices, we found no clear indications that it can be a significant effect for pulsars whose braking indices deviate from the canonical values. This may be due to the fact that the electromagnetic kick theory only considers the electromagnetic emission of stars, and has been developed in the context of a vacuum magnetosphere for the star. Therefore, while the rocket effect predicts braking index values between  3 or 5, the actual values are in many cases considerably different. If we take into consideration the deceleration due to the emission of gravitational waves, the expressions \ref{n1}, \ref{n2} would be very different, and maybe they would end up to more realistic values for the braking indices. Also in this study, it is taken into account that the stellar magnetosphere is vacuum, an assumption that greatly simplifies our calculations but it does not correspond to reality. So perhaps this assumption contributes to the deviation of the results from the real values of the braking indices.  Moreover, it is important to understand the different methodologies that have been used to calculate the values of $n$, that are shown in Table \ref{3.1t}. These results arise from studies that do not only consider dipole emission, but also other conditions. \cite{2020} have results for 19 pulsars based on Parkes radio telescope observations combined with a series of times of arrival (ToAs) simulations. Also \cite{https://doi.org/10.1046/j.1365-8711.1999.02737.x} developed a method to calculate braking indices for intermediate-age stars, without requiring knowledge of ${\ddot{f }}$ (where $f$ is the rotation frequency of the star), while \citet{gourgouliatos2015hall}, took into account the quadrupole components of the magnetic field that appear due to the Hall effect inside the star.

We estimated the kick velocity because of the rocket effect mechanism for 13 neutron stars, initially following the original theory of \cite{Harrison1975AccelerationOP}. We also calculated the kick by developing an alternative model for the rocket effect, in which we included the time dependence of the force, and specifically the age of the SNR and the spin-down of the star. In both cases the velocity was determined for the case where the initial rotation period of the star was 1 $\rm{ms}$ and for the case where we calculate the initial period from equation \ref{tau_c}. The highest kick velocities are found when we apply the theory of Harrison and Tademaru, and consider that $\rm{P_{in}}= 1$ $ms$. For this case the velocity exceeded the 2000 $\rm{km/s}$.Our alternative method gives much lower velocities, whatever the assumption for the initial period of the star. Since we worked mostly on pulsars that are associated with SNRs, maybe this result indicates that perhaps the rocket effect theory should be applied to isolated neutron stars.

In addition, we calculated the displacement that the axis of the stellar magnetic field should have from the center of the star, so that the thrust velocity imparted by the electromagnetic kick mechanism is equal to the current transverse component of the star's spatial velocity. The resulting values of displacement $s$ were within permissible values, although in some cases it was quite large. For example, the highest value for the distance was 9.77 $\rm{km}$ for the pulsar J1809-1917, which is quite extreme. However, in all cases the value of the displacement is more than zero and less than 10 $\rm{km}$, and considering that we assumed a radius of 12 $\rm{km}$ for the neutron stars, the results are optimistic.

To quantify our results, we run a Markov Chain Monte Carlo algorithm for our population of pulsars, to fit our model with the current values of the transverse velocity of each one. By considering the sample, we found that parameters such as the distance of the magnetic axis from the centre of the star and the initial spin period should be $s=7.8$ $\rm{km}$  and $P_{in}=3.8$ $\rm{ms}$ for the rocket effect mechanism to be completely responsible for the velocities we observe today in pulsars. Those results seem reasonable, especially when it comes to the initial spin period which is expected to be quite small. 

In the following sections of this work, we stood on some case studies which are in general considered good candidates to test the rocket effect model. Initially we focused our study on the J0030+0451 pulsar, which even though has a very fast rotation and is verified that its magnetic field consists of multipolar components, seems to not have been though a significant kick, such as the electromagnetic kick, since it has a slow transverse velocity that reaches up to 17 $\rm{km/s}$. For this pulsar we concluded that if the electromagnetic kick had acted on it, its current velocity should exceed 200 $\rm{km/s}$. The estimated values for the distance $s$ an the initial spin period are $s=3.02$ $\rm{km}$  and $P_{in}=3.52$ $\rm{ms}$ respectively. The second case study was the J0538+2817 pulsar. By running again an MCMC algorithm for this neutron star, we found that the values of the distance of the magnetic axis from the star's center and the initial spin period should be $s=2.91$ $\rm{km}$  and $P_{in}=1.77$ $\rm{ms}$ respectively. That may be due to the fact that, even though this pulsar has a slower rotating period than J0030+0451 today, it has a stronger magnetic field and much a higher transverse velocity, which could mean that a kick mechanism has probably acted on this pulsar.

It seems that the rocket effect can describe some particular cases of neutron stars, as long as they have very small initial spin periods and relatively strong magnetic fields. A good indication that the electromagnetic kick or any other kick have appeared on a pulsar at some point, is its spatial velocity. A large value, thus we have a fast pulsar, it is more likely that we have a good candidate to test our kick mechanism. Also, the radial velocity is very important for the study of kick mechanisms, and it can be estimated by using the Doppler shift of the star's spectral lines, that are going through extreme broadening because of the large magnetic fields of neutron stars. So usually it is hard to know a pulsar's radial velocity. Another important observation, is that in the analysis above, we do not take into consideration the emission of gravitational waves, which are factors that could change our results and give more precise and realistic models, if applied.  It is more likely that if the spin down of the pulsar is mostly because of the gravitational wave emission, then the rocket effect mechanism should have a lesser impact. In their study, \cite{2001ApJ...549.1111L}, found an expression for the kick because of the rocket mechanism, including the spin down time due to the electromagnetic radiation and the gravitational wave radiation of the pulsar. They concluded that for the rocket effect to be viable, the timescales of both phenomena should be almost the same. 

 The presence of a plasma-filled magnetosphere may lead to different results compared to the vacuum electromagnetic field we are considering in this work. For instance, a combination of a dipole and a quadrupole magnetic field, whose axes cross the centre of the star, leads to an asymmetric north-south configuration with respect to the magnetic field. In vacuum, both, the dipole and quadrupole term, exert comparable torques \citep{10.1093/mnras/stw2050} and, arguably, accelerations. Thus, they could activate a rocket mechanism. On the contrary, if a plasma-filled magnetosphere surrounds the star, spin-down is related to the magnetic flux crossing the light cylinder and this is mostly due to the dipole term, provided that the light cylinder is sufficiently larger than the stellar radius \citep{2016ApJ...833..258G}. Thus, while the field is asymmetric on the surface, it is symmetric on the light cylinder.  Nevertheless, given that the rocket effect is significant for millisecond pulsars, even in the case of a plasma-filled magnetosphere, the magnetic flux crossing the light cylinder will depend both on the dipole and quadrupole terms and even in this case the rocket effect may operate. Given these complications, a detailed study of such systems is required to resolve these effects.

Additionally, one very important issue, as discussed above, is the determination of the initial spin period of pulsars.  If we make a better estimation of the initial spin period of neutron stars, we will have more accurate and realistic results.  Here we make a series of tests to find the best conditions for the rocket effect to be realistic. First, we relied on the work of \cite{igoshev2022initial} because initially in this work, we studied pulsars that are associated with SNRs. For this case, if the  characteristic age is greater than the age
of the SNR  we assume that the initial rotation period is as low as 2 ms. This assumption gives small values for the kick velocities. For some pulsars it seems that the rocket effect does not operate while in some cases, we have a maximum value of almost $180$ $\rm{km/s}$. An extreme scenario which ends up to greater kick velocities, is considering very small initial spin periods, equal to 1 $\rm{ms}$. We made this calculation to test the extreme values of the kick that can arise from the rocket effect. Finally we made the MCMC analysis to get more realistic expected values for the initial spin period for our population of pulsars.

Finally, in this study, we made the assumption that the current transverse velocity is attributed entirely to only one kick mechanism. However, it is more likely that more than one kick-mechanisms can accelerate the star, for different timescales and in different time periods of the pulsar's life. As mentioned above a main difference between the described mechanisms (see Table \ref{kicks}) is that the rocket effect is a post natal kick while the other kicks appear when the pulsar is born. Additionally the rocket effect is expected to have much greater timescales that could reach up to years while the neutrino driven kick and the hydrodynamical kick are expected to have timescales of a few seconds. With a quick calculation by using equation \ref{10}, someone has to make the integration for almost 10 years to have a kick velocity of  almost 100 $\rm{km/s}$. So it is probably necessary to combine two or more velocity kick mechanisms to understand the kinematics of neutron stars.

\section*{Data Availability}

The data underlying this article will be shared on reasonable request to the corresponding author.

\section*{Acknowledgements}

The authors are very grateful to an anonymous referee for her/his thorough review and comments that improved the manuscript.



\bibliographystyle{mnras}
\bibliography{example} 




\appendix

\section{Multipolar expansion of the electromagnetic fields}

Let us consider a neutron star with a displaced magnetic axis from the star's center, that we symbolise as $s$.  Let us expand this offset dipole filed in multipoles. We assume that the neutron star is rotating in vacuum, thus, electric charges and currents are zero at the magnetosphere. Then we introduce time-dependence in the fields through the following expression, ${\rm e}^{-ikx-i\omega t}$, with ${k=n_f\omega/c}$ where $\omega$ is the angular frequency and $c$ is the speed of light, and $n_f$ indicates the $n_fth$ Fourier component. Therefore, the final  expressions for the electromagnetic fields \citep[Chapter 9]{Jackson:100964} are the following,

\begin{equation}
  \begin{aligned}[b]
    & {\bf H }=\frac{e^{ikr-i\omega t}}{kr} \sum_{m=-l_{max}}^{l_{max} }\sum_{l=0}^{l_{max}} (-i)^{l+1}\\
    &\left[ \alpha_E(l,m){\bf X}_{lm}(\theta, \phi)+   
     \alpha_M(l,m){\bf \hat{ n}} \times {\bf X}_{lm}(\theta, \phi)\right] 
     \end{aligned}
\end{equation}

\begin{equation}
    {\bf E}= Z_0 ({\bf H} \times {\bf \hat{ n}}), \hspace{0.8cm} {\bf \hat{n}}=\frac{\bf r}{r}.
\end{equation}

In equations (2) and (3), $\bf {H, E }$ stand for the magnetic and electric field respectively. These expressions contain the degree of the multipole expansion $(l,m)$ and the constant $\rm{Z_0=\sqrt{{\mu_0}/{\epsilon_0}}}$. The index $l=1$ for a dipole, $l=2$ for quadrupole etc., while $m=0,\pm 1, \pm 2$. The multipole coefficients $\alpha_E(l,m), \alpha_M(l,m)$, give the contribution of each degree of the multipole, and are equal to the following expressions,
\begin{equation}\label{2e}
    \alpha_E(l,m)=\frac{ik^3}{\sqrt{l(l+1)}}\int{Y^*_{lm}(\theta,\phi) j_l(kr){\bf R} \cdot\left({\bf M}+\frac{1}{k^2}{\bf \nabla} \times {\bf J}\right) \,d\omega },
\end{equation}
\begin{equation}\label{3e}
    \alpha_M(l,m)=\frac{-k^2}{\sqrt{l(l+1)}}\int{Y^*_{lm}(\theta,\phi) j_l(kr){\bf R} \cdot({\bf J}+{\bf \nabla} \times {\bf M}) \,d\omega }.
\end{equation}

Moreover the term ${\bf {X}}_{lm}(\theta,\phi)$ is equal to:
\begin{equation}
{\bf{X}}_{lm}(\theta,\phi)= \frac{1}{\sqrt{l(l+1)}} {\bf{R}} Y_{lm}(\theta,\phi).
\end{equation}

In equations \ref{2e} and \ref{3e}, the Bessel functions $j_l(kr)$ appear, and also the spherical harmonic functions $Y_{lm}$. Moreover ${\bf M}$ is the magnetization which has been expanded into a Fourier series, with the nth Fourier component:

\begin{equation}
    {\bf M} ({\bf r},t)= \sum_{n_f=-\infty}^{\infty} {\bf M}_{n_f}({\bf r}) e^{-i n_f \omega t},
\end{equation}
and $\bf J$ is the current density. Also the vector $\bf R$ equals to
\begin{equation}
 {\bf {R}}= \frac{1}{i}({\bf r} \times {\bf \nabla}).   
\end{equation}

Since we know the multipole coefficients we can determine the power of the emission that has been radiated (luminosity) $L$:
\begin{equation}\label{4}
    L=\frac{c}{2\pi} \sum_{n=1}^\infty \frac{1}{k^2}\sum_{l,m}\left[ |\alpha_E(l,m)|^2+|\alpha_M(l,m)|^2\right].
\end{equation}

\cite{Harrison1975AccelerationOP} focused on the case where
$kr \ll 1$ (wave zone or near zone) where the intensity of electromagnetic
fields decreases faster with distance compared to the radiation zone. Keeping the terms with the most important contributions (n=1) and taking into consideration the components of the magnetic dipole moment, the luminosity can finally be written as:

\begin{equation}
    L=\frac{2\omega^4}{3c^3}\left[ \left(\mu_{\rho}^2+\mu_{\phi}^2\right) + \frac{2}{5}\left(\frac{\omega s \mu_z}{c}\right)^2\right].
\end{equation}

This equation can be used for the determination of the reaction force that eventually accelerates the neutron star, which is equal to:
\begin{equation}
    { F}=\frac{2\omega^4}{3c^4}\left[ \left(\mu_{\rho}^2+\mu_{\phi}^2\right) + \frac{2}{5}\left(\frac{\omega s \mu_z}{c}\right)^2\right],
\end{equation}
with $\mu_z, \mu_\rho, \mu_\phi$, the magnetic dipole components in cylindrical coordinates ($\rho, \phi, z$). If we choose the values of $\rm{\mu_\rho \approx 0, \mu_z \approx \mu_\phi}$, we derive to the final expression for the net force due to emission of asymmetric radiation \citep{2001ApJ...549.1111L}:
\begin{equation}\label{F}
    { F}=\frac{8 \omega^5 \mu_z^2 s}{15 c^5}.
\end{equation}

We should also take into consideration the asymmetric factor  $e= cF/L$, which is of order $0.4 \hspace{0.1cm} \omega s / c$ and for a given $\omega$ the maximum value is $e_{max}=0.63$ for $\mu_\rho/\mu_z= \sqrt{0.4} ( \omega s / c)$. Moreover assuming that the star does not emit gravitational but only electromagnetic radiation then the spin-down luminosity \textit{$L=- e {I\omega \dot{\omega}}$}, where $I$ is the moment of inertia of the neutron star. {The kick velocity can be found by combining Newton's second law and the fact that the electromagnetic force is equal to \textit{$ F=Le/c$}:}
\begin{equation}
    F=M \frac{du}{dt}=-e\frac{I\omega \dot{\omega}}{c}.
\end{equation}

{Then the final expression that Harrison and Tademaru arrived to is equation \ref{eq1} as modified by \citep{2001ApJ...549.1111L}.}
In equation \ref{eq1} if we consider an initial spin period of $P_{in}=1$ $\rm{ms}$ and a distance $s=10$ $\rm{km}$ for the magnetic axis from the star's center, we expect a kick velocity that can exceed 500 $\rm{km/s}$ \citep{2001ApJ...549.1111L, 2021}. We note the above expressions for the total spin-down luminosity, force and therefore the velocity of the kick can be evaluated following a different process. Alternatively, by integrating the Poynting flow per solid angle unit, and taking into account the retarded Lienard-Wiechert potential, we can find the
corresponding equations that interest us.

By observing equation \ref{eq1}, it is obvious that the initial spin period of the pulsar is an important parameter for the rocket effect model and its value can affect the final result of the kick velocity. The faster a pulsar rotates, the more likely it is for the rocket kick to appear. In the following sections a discussion about initial spin periods of pulsar is being made.




\bsp	
\label{lastpage}
\end{document}